\documentclass[10pt]{amsart}
\usepackage{amsfonts,amsmath,amsthm,eucal}
\usepackage{epic,eepic,graphicx,amsfonts}
\begin{document}
\title{R{\'e}nyi extrapolation of Shannon entropy }
\author[K. \.Zyczkowski]{Karol \.Zyczkowski
}
\address{Instytut Fizyki im. Smoluchowskiego,
Uniwersytet Jagiello{\'n}ski,
ul. Reymonta 4, 30-059 Krak{\'o}w, Poland\\
and \\
Centrum Fizyki Teoretycznej,
Polska Akademia Nauk, Al. Lotnik{\'o}w 32/44, 02-668 Warszawa, Poland}
\email{karol@tatry.if.uj.edu.pl}

\bigskip
\begin{abstract}
  Relations between Shannon entropy and R{\'e}nyi entropies
  of integer order are discussed.
  For any  $N$--point discrete probability
  distribution for which the R{\'e}nyi entropies of order two and three
  are known,
  we provide an lower and an upper bound for the Shannon entropy.
  The average of both bounds provide an explicit
  extrapolation for this quantity.
  These results imply relations between the von Neumann entropy
  of a mixed quantum state, its linear entropy and traces.

\end{abstract}
\maketitle

\centerline{ver. 2 with corrigendum added, \ \ February 17, 2005}

\section{Introduction}

We are going to analyze discrete probability distributions
${\vec x}=\{x_1,..x_N\}$, which consist of non--negative numbers
summing to unity. $\sum_{i=1}^N x_i=1$.
To characterize quantitatively such probability vectors
one uses Shannon (information) entropy \cite{Sh48}
\begin{equation}
  H({\vec x}):=
    -  \sum_{i=1}^N x_i \ln x_i
 \label{Shann1}
\end{equation}
where we adopt the convention that $0 \ln 0 =0$, if necessary.

The Shannon entropy is distinguished by several unique
properties \cite{Kh57},
 but it is often convenient to introduce generalized 
{\sl R{\'e}nyi entropies} \cite{Re61}
parametrized by a continuous parameter $q$,
\begin{equation}
H_{q}({\vec x}) :=  \frac{1}{1-q}
\ln
\Bigl[ \sum_{i=1}^N x_i^q\Bigr].
\label{PSalph1}
\end{equation}
The R{\'e}nyi entropies are well defined for positive $q \ne 1$,
but is is not difficult to show that
for any probability distribution $\vec x$
one has  $\lim_{q \to 1} H_{q}({\vec x})=H({\vec x})$.
For consistency the Shannon entropy $H$
will thus be denoted by $H_1$.
This method of generalizing the Shannon entropy is
by far not the only one -- for reviews of other generalizations
see books by Kapur \cite{Ka95} and Arndt \cite{Ar01}. 

In this work we discuss relations between
R{\'e}nyi entropies of different orders, and in particular
between $H_1$, $H_2$ and $H_3$. 
Physical motivation for such a study
is twofold. First, we may not know the entire
vector $\vec x$, but only a few of its $L_q$ norms,
so knowing the Renyi entropies, say $H_2$ and $H_3$
we want to estimate the unknown Shannon entropy $H_1$.
Such a situation occurs if one studies an
$N$ dimensional quantum mechanical mixed state $\rho$
according to the scheme recently proposed by
P. Horodecki et al. \cite{HE02,AHOKE03}
and measures directly the traces
Tr$\rho^k$ for $k=2,3,...M$.
If $M<N$ the entire spectrum of $\rho$ remains unknown,
and it is not possible to find its von Neumann entropy,
$S(\rho)=-{\rm Tr}\rho\ln \rho$,
(i.e. the Shannon entropy of the spectrum),
but the generalized Renyi entropies $H_k$,
including the linear entropy, which is a function of $H_2$
may be readily obtained.
Similar problems arise in many different branches of physics,
for instance by the study of the statistics between particles created in
high--energy collisions \cite{BC00, BCO03}.
Measuring probabilities of that two independent 
collisions give rise to the same distribution of particles
allows one to obtain the R{\'e}nyi entropy $H_2$,  
but not directly the Shannon entropy $H_1$.

Another reason to study relations between $H_1$ and $H_2$
has been provided by the work by Pipek and Varga \cite{PV92}.
They assumed that both these quantities are known,
and observed that their difference,
$S_{str}:=H_1-H_2$
called {\sl structural entropy},
provides an important characterization of the
analyzed probability vector $\vec x$.
For instance, an increase of
the structural entropy charactering an eigenstate
of a tight binding model indicates the Anderson transition.
Several other applications of structural entropy
include also quantum chemistry and
statistical analysis of quantum spectra,
(see \cite{VP02} and references therein).

The von Neumann entropy of a
mixed state obtained by partial trace of a
bi--partite pure state,
$\rho={\rm Tr}_B(|\psi\rangle \langle\psi|)$,
measures the degree of entanglement of the
pure state $|\psi\rangle$. Alternatively
one can measure the entanglement by generalized entropies
(see e.g. \cite{ZB02}), so relations between entropies
analyzed in this work provide bounds between different
measures of entanglement.
This very point has recently been discussed
in the paper by Wei et al. \cite{WNGKMV03},
which provides an additional motivation for the present work.

This paper is ogranized as follows. In section 2
the basic properties of the R{\'e}nyi entropies
are reviewed. In section 3 we present
recent results of Tops{\o}e and Harremo{\"e}s \cite{To00,HT01},
which allow us to propose lower and upper bounds  on the Shannon entropy
obtained out of the  R{\'e}nyi entropies of order two and three,
provided the length $N$ of the vector is known.
They are derived in section 4, while
in section 5 we propose and
analyze en estimation of the Shannon entropy.

\section{Shannon and R{\'e}nyi entropies}

Consider a random variable $\xi$ attaining not more
than $N$ different values  with probabilities
$x_i$, $i=1,\dots,N$. Such discrete
probability distribution $P$ may be cahracterized by the
Shannon entropy (\ref{Shann1})
or generalized R{\'e}nyi entropies (\ref{PSalph1}).

All generalized entropies $H_q$ vary from zero for a certain event
(the distribution $Q_1:=\{1,0,...,0\}$) to
$\ln N$, for the uniform distribution,
(the distribution $Q_N:=\{1/N,1/N,...,1/N\}$).
For the distributions with $k$ equal elements,
 $Q_k:=\{1/k,...,1/k,0,...,0\}$, the entropies
admit intermediate values, $\ln k$.

The R{\'e}nyi entropy $H_q$ converges to the Shannon
entropy in the limit $q \to 1$.
It is also useful to express the Shannon entropy as
the limit of the derivative,
\begin{equation}
H({\vec x}) = -\lim_{q\to1}
\frac{ \partial[(1-q)\exp\bigl(H_q({\vec x})\bigr)]}
        {\partial q}.
\label{PSderiv}
\end{equation}

Some special cases of $H_{q}$  are of special
interest. For $q=2$ we have
$H_2(\vec{x}) =-\ln [ \sum_{i=1}^N  x_i^2]$.  The
R{\'e}nyi entropy of order two, called {\sl extension entropy}
 \cite{PV92}, is closely related to the
{\sl inverse participation ratio},
\begin{equation}
R({\vec x}) := \frac{1}{\sum_{i=1}^N  x_i^2 }
= \exp [H_2({\vec x})].
\label{IPR}
\end{equation}
This quantity characterizes
the "effective number of different events"
which the stochastic variable may admit,
and varies from unity for $Q_1$,
to $N$ for the uniform distribution $Q_N$.
Another quantity $r=1/R=\sum_{i=1}^N  x_i^2$,
called {\sl index of coincidence} \cite{HT01},
in quantum mechanical problems is called {\sl purity}, 
since the larger $r$ the more pure, the state it describes.
The quantity $L:=1-r=1-\exp(-H_2)$ is called {\sl linear entropy}
since in analogy to Shannon entropy
it achieves its maximum for the uniform distribution $Q_N$.

In the case $q=0$ the R{\'e}nyi entropy is a function of
the number $m$ of positive components of the vector,
$H_0(\vec{x})=\ln m$.
In the limit $q \to \infty $
we obtain a quantity analogous to the Chebyshev norm:
$H_{\infty}=-\ln x_{\rm max}$, where $x_{\rm max}$ is the largest
component of $\vec x$.

The R{\'e}nyi entropy (\ref{PSalph1}) is a sum of $N$
terms so for any finite $N$
the function of $H_q$ on $q$ is differentiable.
The functional dependence of the R{\'e}nyi entropy
on its parameter was investigated
in detail by Back and Schl{\"o}gl \cite{BS93}).
Making use of the fact that the function $x^s$ is convex for $s>1$ and concave
for $0\le s \le 1$ they have proved
several inequalities\footnote{In the book\cite{BS93}
the quantity $-H_q$ called {\sl R{\'e}nyi information} was analyzed,
so the direction of the inequalities derived there is inverted.},
which we recall in the case $q>0$,
 \begin{eqnarray}
\frac {\partial } {\partial q} H_q \le 0,
 \label{diff1} \\
\frac {\partial } {\partial q}   \frac {q-1}{q} H_q \ge 0,
\label{diff2} \\
\frac {\partial } {\partial q}  (1-q) H_q \le 0, \\
\frac {\partial^2 } {\partial q^2}  (1-q) H_q \ge 0 .
\label{diff4}
\end{eqnarray}

The first inequality (\ref{diff1})
means that the R{\'e}nyi entropy is
a non increasing  function of its parameter,
\begin{equation}
H_{q}({\vec x}) \ge H_{s}({\vec x})
\qquad {\rm for \qquad  any} \qquad
s  > q
\label{Renmonot}
\end{equation}
and this statement is valid also for
infinite probability vectors
and the cases of nondifferentiable $H_q$ \cite{BS93}.
Hence the structural entropy
$S_{str}:=H_1-H_2$  is non--negative \cite{PV92}.

Inequality (\ref{diff4}) implies that
the dependence of the R{\'e}nyi entropy
on its parameter is convex.\footnote{This claim is
withdrawn - see into corrigendum
Sec. \ref{corri}, in which
consequences of this error are pointed out.}
Thus knowing the $0$ and $2$--entropies one
obtains by linear interpolation an
upper bound for the Shannon entropy
\begin{equation}
H_{1}({\vec x}) \le H_{u0}:=\frac{1}{2}(H_{0}({\vec x})+H_{2}({\vec
x}) ).
\label{Ren02}
\end{equation}
This relation gives us an upper bound for the
structural entropy
\begin{equation}
S_{str} ({\vec x}):=H_{1}({\vec x})-H_2({\vec x})
 \le \frac{1}{2} \bigl
(H_{0}({\vec x})-H_{2}({\vec x}) \bigr)\
\label{Ren002}
\end{equation}
valid for any vector ${\vec x}$  of a finite length $N$.

In an analogous way, if the R{\'e}nyi entropies
of order $2$ and $3$ are known, the
linear extrapolation provides a lower bound
for the Shannon entropy
\begin{equation}
H_{1}({\vec x}) \ge H_{d23}({\vec x}) = 2H_{2}({\vec x})-H_{3}({\vec x}) ,
\label{Ren23}
\end{equation}
which combined with (\ref{Ren02}),
allows one to write down a simple estimation
$H_{023}=(H_d+H_{u0})/2$, (see  Fig 3.a),
\begin{equation}
H_{1}({\vec x}) \approx H_{023}({\vec x}):=\frac{1}{4}[H_{0}({\vec
x})+5H_{2}({\vec x}) -2  H_{3}({\vec x}) ] .
\label{Ren023}
\end{equation}

Making use of the inequality (\ref{diff2})
we obtain the relation
\begin{equation}
\frac{q-1}{q} H_{q}({\vec x}) \le
     \frac{s-1}{s}    H_{s}({\vec x})
\qquad {\rm for \quad  any} \qquad
s  \ge  q \ ,
\label{Renilor}
\end{equation}
which is equivalent to the statement
that the $L_q$--norm is a non--increasing
function,  
$||{\vec x}||_s \le ||{\vec x}||_q$.
This result provides another upper bound,
$H_{q}\le q(s-1)H_s/s(q-1)$.
Although it is not applicable
for the Shannon entropy,
for which $q=1$ so the inequality becomes trivial,
but it gives an usefull bound on $H_q$ with $g>1$
by the limiting value $H_{\infty}$,
\begin{equation}
 H_{q}({\vec x}) \le
     \frac{q}{q-1}  H_{\infty}({\vec x}) .
\label{Reniinft}
\end{equation}

In further sections of this work we shall discuss possibilities
of finding more precise
bounds and estimations for the Shannon entropy,
provided the dimension $N$ of the probability vector is known.

\section{Bounds between R{\' e}nyi entropies}

For any value $q \ge 0$ the generalized entropy
$H_{q}$ is equal zero for certain events described by
the distribution $Q_1$, and achieves its
maximum for the uniform distribution,
$S(Q_N)=\ln N$.

To investigate further relations between the R{\' e}nyi entropies of different order
we have chosen to analyze the case of $N=3$ dimensional vectors $\vec x$.
The space of all possible probability vectors, plotted in the
the plane $x_3=1-x_1-x_2$ forms an equilateral triangle
of side $\sqrt{2}$ measured in the Euclidean distance.
Its three corners: $(100)$, $(010)$ and $(001)$
represent certain events, while the center of the triangle
corresponds to the uniform distribution $Q_3$.

\begin{figure} [htbp]
   \begin{center}
\
 \vskip -0.2cm
 \includegraphics[width=9.5cm,angle=0]{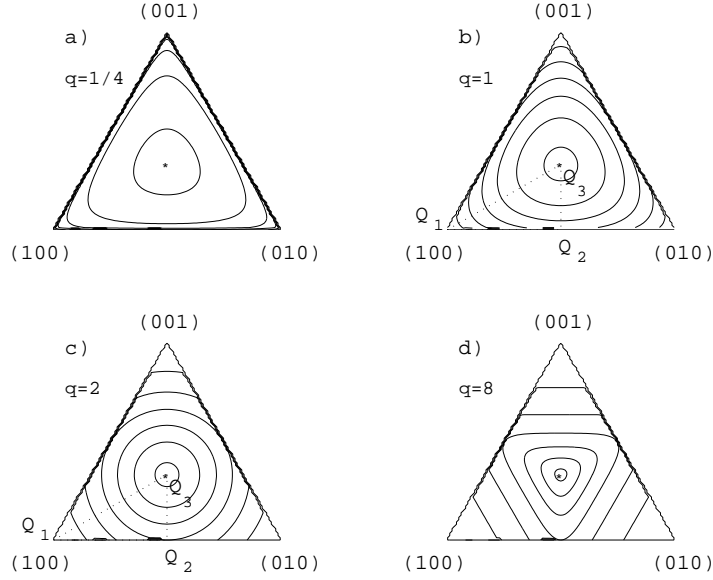}
\caption{
 Iso-entropy curves in the space of probability distributions
with  $N=3$. The generalized R{\'e}nyi entropy
is constant along the curves plotted for
(a) $q=1/4$,
(b) $q=1$ (Shannon entropy),
(c) $q=2$ (Euclidean circles--distance $D_2$), and
(d) $q=8$. Dotted lines form the triangle $\triangle (Q_1,Q_2,Q_3)$.
}
 \label{fig:ren1}
\end{center}
 \end{figure}

Fig.  \ref{fig:ren1} shows sets of points
characterized by the same R{\'e}nyi entropy of order $q$,
which may be called iso-entropy curves.
Independently of the value of the parameter $q$
the generalized entropy attains its minimum, $H_{q}=0$,
at the corners of the triangle, while
the maximum $H_{q}=\ln 3$ is achieved
at the point $Q_3$ at the center of the triangle.
As shown in Fig. \ref{fig:ren1}a
the maximum is rather flat for $q=1/4$. The case shown in
this panel resembles the limiting case $H_0$, for which
the entropy reflects the number of events which may occur:
it vanish at the corners of the triangle, is equal to
$\ln 2$ at its sides and equals to $\ln 3$ for any point
inside the triangle.
The other example, $q=8$, presented in Fig.
 \ref{fig:ren1}d.
is similar to the limiting case $H_{\infty}$,
for which the iso-entropy curves are perpendicular to the lines joining
$Q_3$ with the corners.

Superimposing some of the above pictures on one graph allows one to
understand further relations between the R{\'e}nyi entropies.
The generalized entropies are correlated; e.g.
for the distributions $Q_k$ the entropies are equal
to $\ln k$ independently of the value of $q$.

The problem, which values the entropy $H_{q}$ may admit,
provided $H_{s}$ is given,
has been solved by Harremo{\"e}s
and Tops{\o}e \cite{HT01}. For any distribution $P \in {\mathbb R}^N$
they proved a simple
(but not very sharp) upper bound on  $H_1$ by $H_2$,
\begin{equation}
H_{2}({\vec x}) \le H_1({\vec x}) \le \ln N+ 1/N -
\exp(-H_2({\vec x}))
\label{PS02inf}
\end{equation}
where the lower bound is a special case of (\ref{Renmonot}).
Moreover, they showed that
the set $\Delta_{q,s}$ of possible probability distributions
plotted in the plane $H_{q}$ versus $H_{s}$
is not convex (see Fig. 2),
and its boundaries are formed of arcs
corresponding to the interpolating probability distributions
\begin{equation}
Q_{k,l}(a):=aQ_k+(1-a)Q_l \qquad {\rm with} \qquad
     a\in [0,1] .
\label{Reninterp}
\end{equation}
More precisely, for any probability distribution $P$
consisting of $N$ components and arbitrary
$s > q >0 $ the following bounds hold \cite{HT01}
\begin{equation}
 H_{q}(Q_{k-1,k}(a))
   \le H_{q}(P)  \le H_{q}(Q_{1,N}(a)) ,
\label{Renibound}
\end{equation}
where $a$ is a function of the known value of
$ H_{s}(P)$ and
 the natural number $k$ is selected by the inequality
$ \ln (k-1) \le  H_{s}(P)  \le \ln k$.

The above results, crucial for the main body of this work,
are easy to understand. Let us discuss the simplest
nontrivial case with $N=3$. The two dimensional simplex of probability distributions
may be divided into $6$ identical parts, equivalent to the
triangle $\triangle (Q_1,Q_2,Q_3)$, as shown in Fig. 1c. 
Three sides of the triangle are formed of the interpolating distributions
$Q_{1,2}$, $Q_{1,3}$ and $Q_{2,3}$ and these distinguished probabability 
distributions are extreme in a sense that 
they lead to the bounds (\ref{Renibound}).
The bounds between $H_1$ and $H_2$  for $N=3$ are presented
in Fig. 2a. To obtain them it is sufficient
to travel along the sides of the triangle $\triangle (Q_1,Q_2,Q_3)$,
computing $H_1$ and $H_2$ at each point and to
plot the data obtained in the plane $H_1$ versus $H_2$. 

More formally, 
the upper boundary of the set $\Delta_{q,s}$
consist of one arc derived from the family 
of distributions $Q_{1,N}(a)$;
for any value of $a$ we compute $H_{s}(a)$,
invert it to obtain $a(H_{s})$
and plot $H_{q}(a(H_{s}))$.
In the case $N=3$ the upper bounds
plotted in Fig. 2a, c and e
arise from the hypotenuse $Q_1,Q_3$
of the triangle $\triangle (Q_1,Q_2,Q_3)$ from 
Fig. \ref{fig:ren1}.b and c.

In the similar way the lower bound may be derived
from the distributions $Q_{k,k+1}(a)$ for $k=1,...N-1$.
It consists of $(N-1)$ arcs forming an cascade
$ H_{q}(Q_{1,2}(a))\smile
 H_{q}(Q_{2,3}(a))\smile \cdots \smile
 H_{q}(Q_{N-1,N}(a))$.
Note that the distributions $Q_k$
are represented in each plot by the points
$( \ln k,\ln k )$, which connect
the neighbouring arcs.
For $N=3$ the lower bound consists of two arcs,
corresponding to the adjacent sides 
$Q_1,Q_2$ and $Q_2,Q_3$ of $\triangle (Q_1,Q_2,Q_3)$.

The shape of the set $\Delta_{q,s}$
requires a comment. The $N-1$ dimensional simplex
-- the set of all $N$--points probability distributions
 is convex and any of its projections onto a plane
forms a convex set. However, its image at the plane
$H_{q}$ versus $H_{s}$ needs not to be convex, since
the transformations
$H_{q}({\vec x})$ and $ H_{s}({\vec x})$ are nonlinear.
The boundaries of
 $\Delta_{q,s}$ are obtained as the image of an appropriately
chosen path on the boundary of the simplex. In the case considered
it is the path $Q_1\to Q_2\to \cdots \to Q_N \to Q_1$, independently of
the values of $q$ and $s$. Observe that the general structure of the
set $\Delta_{q,s}$ does not depend on $s$. However, the  larger difference $s-q$,
the larger area of the set: the less information on $H_q$ is provided by $H_s$.

\begin{figure} [htbp]
   \begin{center}
\
 \vskip -0.2cm
 \includegraphics[width=9.5cm,angle=0]{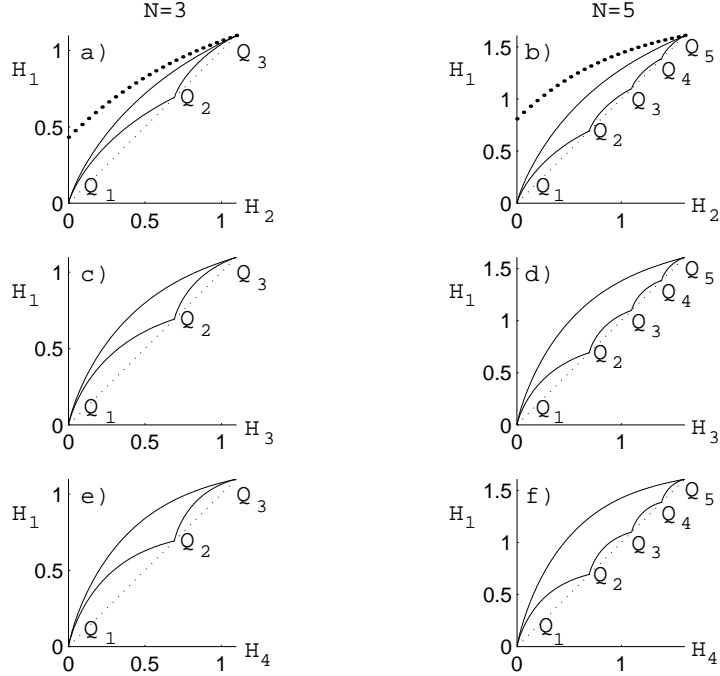}
\caption{
The set of all possible discrete distributions
for $N=3$ and $N=5$ at the R{\'e}nyi entropies
plane  $H_1$ and $H_{q}$:
 $q=2$ (a,b);
 $q=3$ (c,d) and
 $q=4$ (e,f).
Thin dotted lines in each panel
represent the monotonicity lower bounds (\ref{Renmonot})
while bold dotted curves in panel (a) and (b)
denote upper bound (\ref{PS02inf}).
}
 \label{fig:ren2}
\end{center}
 \end{figure}

Let us emphasize in this point that results 
presented in \cite{HT01}
do not close the issue of finding bounds and relations 
between different entropies.
Results analogous to (\ref{Renibound})
for a more general class of entropy functions were recently obtained by 
Berry and Sanders \cite{BS03}.
A more precise lower bound for Shannon
entropy, quadaratic in terms of 
index of coincidence (purity) was found by Tops{\o}e \cite{To03}.

\section{$N$--dependent bounds for Shannon entropy}
\subsection{Bounds based on  $H_2$ and the length $N$ of the probability vector}

Results (\ref{Renibound}) allow us to obtain
bounds for a value of the entropy $H_{q}$,
provided the value $H_{s}$ is known.
Let us first assume that the entropy $H_2$ is known and we want to
extract some information on $H_1$.
We start computing the R{\'e}nyi entropy of order two,
\begin{equation}
 H_{2}(Q_{1,N}(a)) = - \ln \Bigl[
 \frac{(1+(N-1)a)^2}{N^2} +(N-1) \frac{(1-a)^2}{N^2}
      \Bigr] \
\label{s2aaa}
\end{equation}
and invert it to obtain
\begin{equation}
 a=\sqrt{ \frac{N\exp(-H_2)- 1}{N-1}  }  .
\label{aaa12}
\end{equation}

In this way we receive sharp upper bounds
for $q \in (0,2)$
\begin{equation}
 H_{q}(P) \le
 \frac{1}{1-q} \ln \Bigl[
  \Bigl( \frac{1+(N-1)a}{N}\Bigr)^{q} +
  (N-1) \Bigl( \frac{1-a}{N}\Bigr)^{q}
   \Bigr] ,
\label{Interaa2}
\end{equation}
 which  for $q \to 1$ reduces to
\begin{equation}
 H_{1}(P) \le H_{12}^{u}:=(1-N) \frac{1-a}{N}\ln \frac{1-a}{N}-
           \frac{1+a(N-1)}{N}\ln \frac{1+a(N-1)}{N} \ ,
\label{Inter12u}
\end{equation}
with $a$ given by (\ref{aaa12}).

To obtain analogous lower bound
we find $k$ such that
$ \ln (k-1) \le  H_{2}  \le \ln k $ and compute
$ H_{2}(Q_{k-1,k}(a))$. Also this relation
may be easily inverted
providing $a={\sqrt{ k(k-1)\exp(-H_2)+1-k}}$.
Thus we arrive at a lower bound for the R{\'e}nyi entropy
\begin{equation}
 H_{q}(P) \ge    \frac{1}{1-q}
   \ln \Bigl[ (k-1)z^{q} + y^{q}
     \Bigl] \ ,
\label{Inter12d}
\end{equation}
and in particular case, for the Shannon entropy
\begin{equation}
 H_{1}(P) \ge H_{12}^{d}:=
      (1-k)z\ln z - y\ln y \ ,
\label{shann12d}
\end{equation}
with $z=(k+a-1)/(k^2-k)$ and $y=(1-a)/k$.

\subsection{Bounds based on $H_3$ and $N$}
Let us now assume, we know the value
of the R{\'e}nyi entropy $H_3$.
As in (\ref{s2aaa})
we compute $H_{3}(Q_{1,N}(a))$,
and invert it finding
$a\in [0,1]$ as the largest (real) root of the
polynomial
\begin{equation}
 W_u(a)=a^3+ a^2 \frac{ 3}{N-2}-
 \frac{N^2\exp(-2H_3)-1}{(N-1)(N-2)} =0.
\label{Interaa3}
\end{equation}
Then the upper bound valid for $q \in (0,3)$
is given by the same formula (\ref{Interaa2})
with $a$ given by the root of
(\ref{Interaa3}) instead of (\ref{aaa12}).
For $q\to 1$ one obtains then the
upper bound for the Shannon entropy
\begin{equation}
 H_{1}(P) \le H_{13}^{u}:=(1-N) \frac{1-a}{N}\ln \frac{1-a}{N}-
           \frac{1+a(N-1)}{N}\ln \frac{1+a(N-1)}{N} \ ,
\label{Inter13u}
\end{equation}
with $a$ determined by (\ref{Interaa3}).

To get the lower bound
we look for $k'$ such that
$ \ln (k'-1) \le  H_{3}  \le \ln k'$. The relation
$ H_{3}(Q_{k'-1,k'}(a))$ may be inverted  explicitly for $k'=2$
providing $a={\sqrt{ [4\exp(-2H_3)-1]/3}}$. For $k'>2$
$a$ is given by the only root of the polynomial
\begin{equation}
 W_d(a)=a^3+  a^2  \frac{3(k'-1)}{2-k'}+
 \frac{(k'-1)^2}{2-k'}[1-{k'}^2\exp(-2H_3)] =0.
\label{Interaa4}
\end{equation}
in the interval $[0,1]$.
The lower bound for the R{\'e}nyi entropies
has the same form as (\ref{Inter12d}) and gives
for the Shannon entropy
\begin{equation}
 H_{1}(P) \ge H_{13}^{d}:=
      (1-k')z'\ln z' - y'\ln y'
\label{shann13d}
\end{equation}
with $z'=(k'+a-1)/(k'(k'-1))$ and $y'=(1-a)/k'$.

\section{Combined extrapolation}

In previous section we obtained two
upper bounds for the Shannon entropy:
(\ref{Inter12d})
stemming from the R{\'e}nyi entropy $H_2$, and
(\ref{Ren23}) obtained from $H_3$.
The latter is in general a worse one\footnote{Knowing
the function $H(q)$ at $q=3$
we have less information on $H_1$, than knowing it
at $q=2$.}, but it allows for a linear extrapolation, which
gives
\begin{equation}
H_{up}({\vec x}):=2H^u_{12}({\vec x})-H^u_{13}({\vec x}).
\label{up23}
\end{equation}
Our numerical results allow us to advance the following

{\bf Conjecture.}

{\sl For any probability distribution ${\vec x}$
the bound
\begin{equation}
H_{1}({\vec x}) \le H_{up}({\vec x})
\label{conjec}
\end{equation}
holds.}

In the same way one may try to extrapolate lower bounds defining
$H_{d}({\vec x}):=2H^d_{12}({\vec x})-H^d_{13}({\vec x})$.
For certain probability vectors this quantity may give
a useful approximation for the Shannon entropy. Interestingly,
a relation analogous to (\ref{conjec}),
$H_{1}({\vec x}) \ge H_{d}({\vec x})$ is not true:
it is violated e.g. if $k\ne k'$.

Making use of the rigorous bound (\ref{Ren23})
and the conjecture (\ref{conjec})
we may suggest to estimate the unknown value of the
Shannon entropy by the mean value
\begin{equation}
H_{*}'({\vec x}):= \frac{1}{2} [H_{up}({\vec x})+
H_{d23}({\vec x}) ] =
H^u_{12}({\vec x})+H_{2}(\vec{x})-
 \frac{1}{2} [H^u_{13}({\vec x})+H_{3}(\vec{x})] ,
\label{estim1}
\end{equation}
which is obtained by combined methods based as well on the
lower as well as on the upper bounds.
Observe that this explicitly computable quantity involves only
R{\'e}nyi entropies $H_2$ and $H_3$ and
the dimension $N$.
This estimation may be improved noting that the rigorous lower bounds
(\ref{Ren23}) and (\ref{shann12d})
are not equivalent. Since for some distributions
the latter bound gives better (higher) results, we may improve
(\ref{estim1}) writing
\begin{equation}
H_{*}({\vec x}):= \frac{1}{2} H_{u}({\vec x})+
\frac{1}{2} {\rm max} \{ H_{d23}({\vec x}), H^d_{12}({\vec x}) \}.
\label{estim2}
\end{equation}
If the value of the zero-entropy, $H_0$, is known,
one may replace the upper bound $H_{u}({\vec x})$ used above by
the minimum, ${\rm min} \{ H_{u0}({\vec x}), H_{up}({\vec x}) \}$.

\begin{figure} [htbp]
   \begin{center}
\
 \vskip -0.2cm
 \includegraphics[width=9.5cm,angle=0]{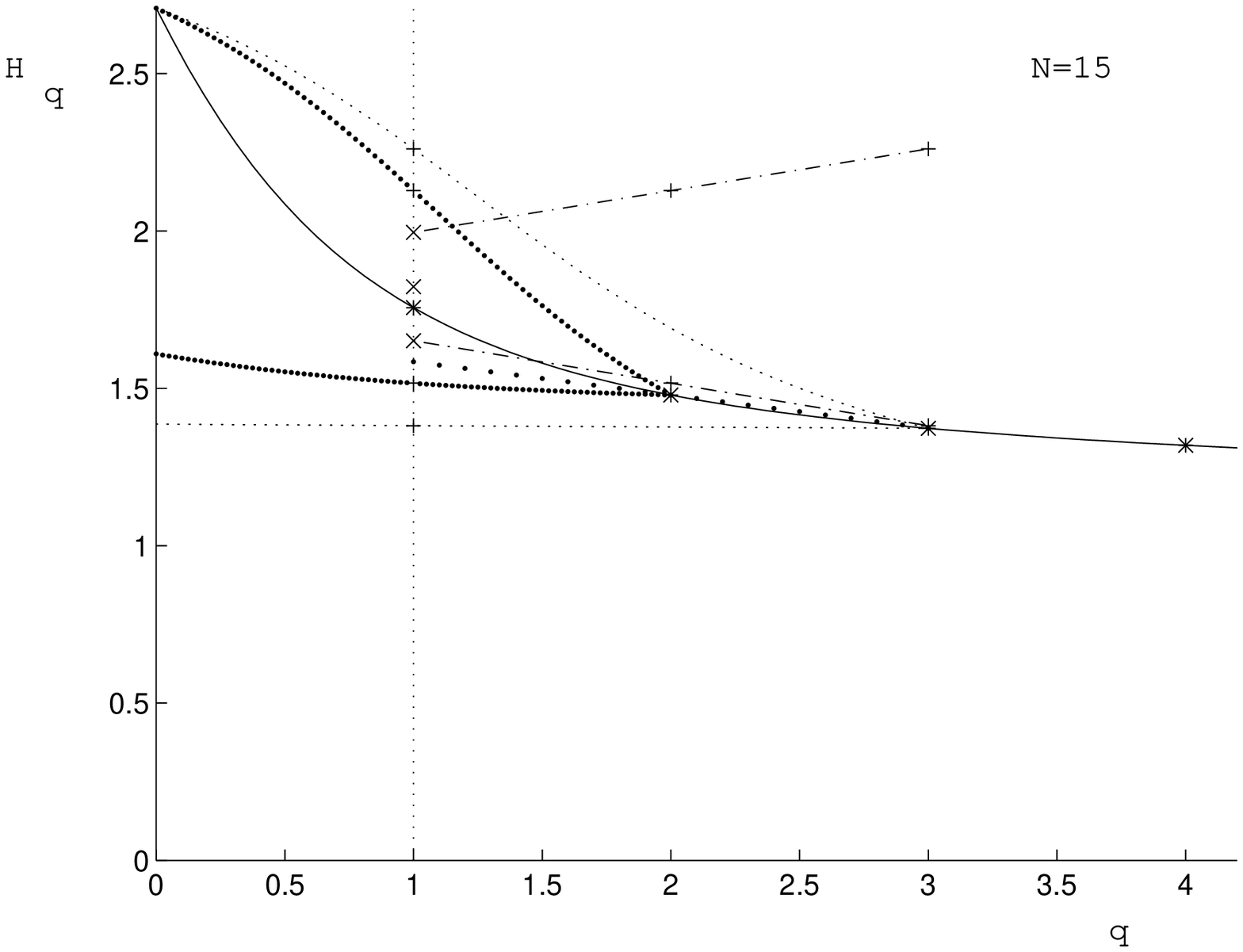}
\caption{
R{\'e}nyi entropies $H_{q}$ for a random probability distribution
  of size $N=15$ (solid line). Dense dotted curves represent
 $H_2$ bounds: upper (\ref{Interaa2}) and lower  (\ref{Inter12d}),
 while faint dotted curve denote analogous bounds based on $H_3$.
 Straight dashed--dotted lines show lower (\ref{Ren23}) and upper
(\ref{up23}) extrapolations.  The exact value of $H_1$ is denoted by $(*)$,
while the estimation (\ref{estim2}) by the middle $(\times )$.
}
 \label{fig:ren3}
\end{center}
 \end{figure}

Figure 3 shows the bounds and the estimations described above for
a randomly chosen probability distributions with $N=15$ components.
The overall quality of the proposed estimations
for the Shannon entropy 
may be judged from Fig. 4, which shows the histogram
of the deviations $\delta_1=H_*-H_1$ and
      $\delta_2=H_d-H_1$ for a sample of $10^4$
probability vectors generated randomly
according to the statistical (Fisher--Rao) measure on the $N-1$ dimensional
simplex \cite{Fi25}. This measure has a simple geometric interpretation:
is suffices to consider a unit
vector ${\vec t}$ distributed uniformly on the sphere $S^{N-1}$,
and to define the probability vector
${\vec x}=\{t_1^2,...,t_N^2 \}$.
Numerical results obtained in this way allow us to conclude
that the proposed estimation  (\ref{estim2})
provides a useful all-purpose approximation 
of the Shannon entropy. Note that the precision of this 
approximation decreases with the length $N$
of the probability vector.

To judge about possible  application of
the estimate $H_*$ in the analysis of physical data, one
should perform analogous numerical simulations
with random vectors ${\vec x}$ generated according to a
specific probability distribution adjusted to a
given physical problem.

\begin{figure} [htbp]
   \begin{center}
\
 \vskip -0.2cm
\includegraphics[width=9.5cm,angle=0]{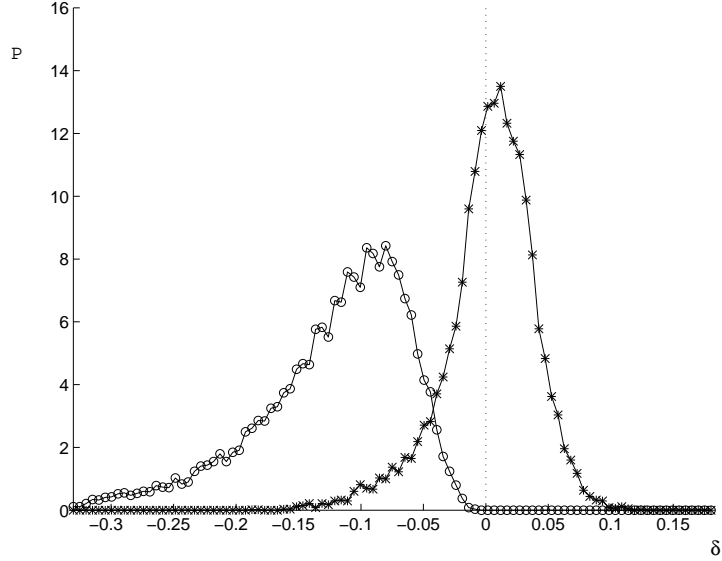}
\caption{Histogram of deviations of the extrapolations
form the real value of the Shannon entropy $H_1$:
 $(*)$ denotes the probability density 
 of error $\delta_1=H_*-H_1$ of the estimation (\ref{estim2}),
  while $(o)$ denotes results for the error
     $\delta_2=H_d-H_1$ of the lower bound (\ref{Ren23})
  for a sample of $10^4$
 random probability vectors of size $N=10$.
 }
 \label{fig:ren4}
\end{center}
 \end{figure}

\section{Concluding Remarks}

In this work we considered the problem of finding the bounds
and extrapolations for the Shannon (entropy), provided
some of the Renyi entropies are known. In general,
generalized entropies of integer order
$q=2$ and $q=3$ are easiest to calculate, and they
are sufficient to obtain bounds
(\ref{Ren02}) and (\ref{Ren23})  for the Shannon entropy.
The quality of the bound may be improved if the
length $N$ of the probability vector is known.
Then an explicit extrapolation (\ref{estim2})
allows us to estimate the actual value of 
the entropy $H$.

Note that the bounds and extrapolations discussed may be easily 
rewritten in terms of a non-extensive entropy
\begin{equation}
S_q(\vec x ) := \frac{1}{1-q}
\Bigl[ \sum_{i=1}^n x_i^q -1 \Bigr]\
\label{PHCalpha}
\end{equation}
used by 
Havrda and Charvat  \cite{HC67} and 
Daroczy \cite{Da70}, which became often used 
in statistical physics after the seminal work of Tsallis \cite{Ts88}.
In particular, the linear entropy $L$
is just the nonextensive entropy of order two,
$L({\vec x}) =S_2(\vec x)$, 
and the bounds between $H_q$ and $H_1$
imply analogous relations between $S_q$ and $H_1$.
In fact the plot presented in \cite{WNGKMV03} 
shows bounds between von Neumann entropy
and the linear entropy and they 
follow directly from relation (\ref{Renibound}) proven in \cite{HT01}.

The issue of comparing the R{\'e}nyi entropies
$H_0$, $H_1$ and $H_2$ is closely related to the 
problem of describing the degree of chaos of an
analyzed classical dynamical system by the 
topological entropy $K_0$, the Kolmogorov--Sinai 
(metric) entropy $K_1$  and the correlation entropy
$K_2$. These dynamical entropies are defined as the 
{\sl rate} of the increase of the 
R{\'e}nyi entropies in time \cite{BS93}, but 
since the length of the probability vector
is not finite the $N$--dependent bounds discussed
in this work are not applicable.
The same concerns comparison of generalized
fractal dimensions $D_q$ of fractal measures:
the box--counting dimension $D_0$,
the information dimension $D_1$ and
the correlation dimension $D_2$,
which also form a non--increasing 
function of the R{\'e}nyi parameter $q$ \cite{BS93},
are defined by the limit $N\to \infty$.

The generalized entropies may also be used to characterize
localization properties of continuous probability distributions.
For instance, any pure quantum state may be represented in the phase
space by the Husimi distribution. Its localization can be measured by
the {\sl Wehrl entropy} defined as the continuous (Boltzmann--Gibbs)
entropy of the Husimi distribution \cite{We78}.
 In an analogous way one may define generalized
R{\'e}nyi--Wehrl entropies \cite{GZ01,VP02}, and
above results may be used to obtain similar bounds
for the Wehrl entropy.

\bigskip

It is a pleasure to thank P. Harremo{\"e}s and F. Tops{\o}e 
for explaining us their results
prior to publication and several fruitful comments.
I am also thankful to I. Bengtsson, A.Bia{\l}as, A. Ostruszka,
F. Mintert, W. Munro and W. S{\l}omczy{\'n}ski
for inspiring discussions and D. Berry, B. Sanders and I. Varga 
for helpful correspondence.
Financial support by Komitet Bada{\'n} Naukowych in Warsaw under
the grant 2P03B-072~19 and by a research grant of the
Volkswagen Stiftung is gratefully acknowledged.

\section{Corrigendum \ \  -- \ \ {\sl February 17, 2005}}
\label{corri}

Eq. (\ref{diff4}) implies
that $(1-q)H_q$ is a convex function of $q$.
However, it does not imply  that 
the R{\'e}nyi entropy $H_q$ 
is a convex function of $q$.
For instance, the R{\'e}nyi entropy 
for a $N=20$ probability  vector
$P=\frac{1}{100}(43,3,\dots,3)$
is not a convex function of $q$.
I am deeply obliged to Christian Schaffner 
for drawing my attention to this fact.

Therefore, equations
(\ref{Ren02}), (\ref{Ren002}), and
(\ref{Ren23}) are not satisfied
and conjecture (\ref{conjec}) cannot hold. 
Furthermore, equations (\ref{Ren023}),
(\ref{estim1}), (\ref{estim2})
may only be used to extrapolate
the unknown value of the Shannon entropy,
from the available data on $H_2$ and  $H_3$.

On the other hand, we would like to mention that the 
lack of convexity of $H_q$ in $q$
does not influence the bounds
between particular values of 
the R{\'e}nyi entropy presented in 
sections 3 and 4 of the present paper.

\bibliographystyle{amsplain}

\end{document}